\documentclass{article}
\usepackage{amsmath}
\usepackage{graphicx}

\title{Duality between magnetic field and rotation}
\author{V. Dzhunushaliev
\thanks{E-mail: dzhun@hotmail.kg}}

\begin{document}
\maketitle

\begin{center}
\textit{Dept. Phys. and Microel. Engineer., KRSU, Bishkek, \\
Kievskaya Str. 44, 720021, Kyrgyz Republic}
\end{center}

\begin{abstract}
It is shown that in 5D Kaluza-Klein theory there are everywhere regular 
wormhole-like solutions in which the magnetic field at the center is the origin 
of a rotation on the peripheral part of these solutions. The time on the 
peripheral part  is topologically non-trivial and magnetic field is suppressed 
in comparison with the electric one. 
\end{abstract}

\section{Introduction}

In this notice we would like to show that in 5D gravitational flux tube solutions 
\cite{dzhsin1} the magnetic field at the central part of the spacetime (throat) can be 
an origin for a rotation at the peripheral parts (tails) of the spacetime. Mathematically 
this duality is based on the fact that for the gravitational flux tube solutions 
there is an interchange of the signs of metric signature on the throat 
$\{+,-,-,-,-\}$ and on the tails $\{-,-,-,-,+\}$ (the solution is regular everywhere). 
It is like to 
Schwarzschield solution in Schwarzschield's coordinates: where the time and radial 
coordinates are interchanged under the event horizon. But there is the 
essential difference: the gravitational flux tube metric is regular everywhere and 
even on the hypersurface where $ds^2 = 0$ the coordinate singularities are missing 
in the contrast with Schwarzschield coordinates which have the coordinate singularities 
on the event horizon. Once this interchanging takes place the time becomes as 
the space-like coordinate and the $5^{th}$ coordinate as the time-like coordinate. 
Then we have to change the metric form in such a way that to isolate the term 
$\left( dx^5_{new} + A_\mu dx^\mu \right)$. Such algebraical manipulations causes 
to the appearance of a new off-diagonal metric term (at the peripheral parts 
of the gravitational flux tube spacetime) $dt_{new} d\varphi$ that is an 
indication of a rotation. Another exhibition of this rotation is that the 
angular momentum density appears for the electromagnetic field. 

\section{Recalculation of electromagnetic fields}

Let us write the 5D metric in the following form
\begin{equation}
	ds^2 = - \delta e^{2\phi} 
	\left(
		d \chi + A_\mu dx^\mu
	\right)^2 + 
	\frac{1}{\delta}dt^2 + 2g_{0i} dt dx^i + 
	g_{ij} dx^i dx^j .
\label{sec0-5}
\end{equation}	
where $A_\mu$ is the electromagnetic potential, 
$x^\mu = t,x^i$ is the 4D coordinates, $\chi$ is the $5^{th}$ coordinate, 
$\mu=0,1,2,3$; $i=1,2,3$. We suppose that exists such everywhere regular 
solution of 5D Einstein's equations that $\delta > 0$ for one part of a spacetime 
(on the throat) and $\delta < 0$ for another one (on the tails). The metric \eqref{sec0-5} 
is written for $\delta > 0$ part. Now we would like to rewrite this metric for the $\delta < 0$ 
part of a spacetime (on the tails). After some algebraical manipulations 
\begin{equation}
	ds^2 =- \widetilde{\delta} e^{2\phi} 
	\left(
	dt + \widetilde{A}_\mu d\widetilde{x}^\mu
	\right)^2 + 
	\frac{1}{\widetilde{\delta}} 
	\left(
	  d \chi + A_i dx^i
	\right)^2 + 
	g_{ij} dx^i dx^j
\label{sec0-10}
\end{equation}
here for the simplicity we consider the case with $g_{0i} = 0$ and 
\begin{eqnarray}
	\widetilde{x}^\mu &=& \left\{ \chi, x^i \right\}; \quad \widetilde{x}^0 = \chi ;
	\quad \widetilde{x}^5 = t,
\label{sec0-20}\\
	\widetilde{\delta} &=& A_0^2 \delta - \frac{e^{-2\phi}}{\delta},
\label{sec0-30}\\
	\widetilde{A}_0 &=& \frac{\delta}{\widetilde{\delta}}A_0
\label{sec0-35}\\
	\widetilde{A}_i &=& 
	\frac{A_0 \delta}{A_0^2 \delta - \frac{e^{-2\phi}}{\delta}} A_i = 
	\frac{A_0 \delta}{\widetilde{\delta}}A_i 
\label{sec0-40}
\end{eqnarray}
where $\widetilde{A}_\mu$ is the new electromagnetic potential (on the tails);
$\widetilde{x}^\mu$ and $\widetilde{x}^5$ are the new coordinates. We suppose that 
$\widetilde{\delta} > 0$ (below we will present solutions with these properties). 
On the peripheral part $t$ is the $5^{th}$ space-like coordinate and $\chi$ is 
the time-like coordinate. 

\section{Application for the gravitational flux tube solutions}

The 5D metric for the gravitational flux tube metric has the following 
form \cite{dzhsin1}
\begin{equation}
  ds^2 = \frac{a(r)}{\Delta (r)} dt^2 - dr^2 - a(r) 
  \left(
  d\theta^2 + \sin^2 \theta d\varphi^2 
  \right) - 
  \frac{\Delta (r)}{a(r)} e^{2\phi (r)} 
  \left(
  d\chi + \omega (r) dt + Q \cos \theta d\varphi
  \right)^2 
\label{sec1:10}
\end{equation}
the functions $a(r), \delta(r) = \Delta(r)/a(r)$ and $\phi(r)$ are the even functions; 
$r \in \{ -\infty , +\infty \}$ and 
consequently the metric has a wormhole-like form; 
$Q$ is the magnetic charge. The off-diagonal metric components are 
$G_{5\mu}/G_{55}=(\omega(r), 0,0, Q\cos \theta )$ and consequently we have  
the radial electric and magnetic fields. The 5D vacuum Einstein's equations are 
\begin{equation}
  R_{AB} - \frac{1}{2} \eta_{AB} R = 0
\label{sec1:20}
\end{equation}
here $A,B$ are 5-bein indices; $R_{AB}$ and $R$ are 5D Ricci tensor and the 
scalar curvature respectively; $\eta_{AB} = diag\{ 1,-1,-1,-1,-1 \}$. 
The 5D substitution of the metric \eqref{sec1:10} in equations \eqref{sec1:20} 
gives us 
\begin{eqnarray}
  \omega'' + \omega'
  \left(
  -\frac{a'}{a} + 2\frac{\Delta'}{\Delta} + 3 \phi '
  \right) &=& 0 ,
\label{sec1:22}\\
  \frac{a''}{a} +\frac{a'\phi'}{a} -\frac{2}{a} + 
  \frac{Q^2 \Delta e^{2\phi}}{a^3} &=& 0 ,
\label{sec1:24}\\
  \phi'' + {\phi'}^2 + \frac{a' \phi'}{a} - 
  \frac{Q^2 \Delta e^{2\phi}}{2a^3} &=& 0 ,
\label{sec1:26}\\
  \frac{\Delta''}{\Delta} - \frac{\Delta' a'}{\Delta a} + 
  3\frac{\Delta' \phi'}{\Delta} + \frac{2}{a} - 6 \frac{a' \phi'}{a} &=& 0 ,
\label{sec1:27}\\
  \frac{{\Delta'}^2}{\Delta^2} + \frac{4}{a} - 
  \frac{q^2 e^{-4\phi}}{\Delta^2} - 
  \frac{Q^2 \Delta e^{2\phi}}{a^3} - 6\frac{a' \phi'}{a} - 
  2\frac{\Delta' a'}{\Delta a} + 2\frac{\Delta ' \phi '}{\Delta}&=& 0 .
\label{sec1:28} 
\end{eqnarray} 
From the Maxwell's equation \eqref{sec1:22} we have 
\begin{equation}
  \omega' = \frac{q a e^{-3\phi}}{\Delta^2} 
\label{sec1:30}
\end{equation}
here $q$ is the electric charge. The solutions of equations \eqref{sec1:22}-\eqref{sec1:28}
are parametrized by electric $q$ and magnetic $Q$ charges \cite{dzhsin1}: 
\begin{enumerate} 
\item 
$0 < Q < q$. The solution is a wormhole-like object. The throat between
the surfaces at $\pm r_H$ \footnote{$r_H$ is defined as follows: 
$\Delta(\pm r_H) = 0$} is \textit{a finite flux tube}
filled with both electric and magnetic fields. 
The longitudinal distance between the $\pm r_H$ surfaces increases by 
$q \rightarrow Q$. We will call these solutions as \textit{gravitational flux tube solutions}. 
\item 
$q = Q$. In this case the solution is \textit{an infinite flux tube} filled
with constant electrical and magnetic fields. The cross-sectional size of 
this solution is constant ($ a= const.$). 
\item 
$0 < q < Q$. In this case we have \textit{a singular finite flux tub}e located 
between two (+) and (-) electrical and magnetic charges located at $\pm r_0$. 
\end{enumerate} 
In this notice we consider the first case. The detailed numerical and approximate 
analytical investigations of the properties of the gravitational flux tube 
solutions is given in Ref. \cite{dzhsin1, dzh2}. This spacetime can be divided on three parts: 
the first one (throat) is the central part of the solution located between 
$r = \pm r_H$ and $\Delta > 0$; 
the second and third parts are located accordingly by $r < -r_H$ and 
$r > r_H$, here $\Delta < 0$. On the throat the electromagnetic potential is
\begin{eqnarray}
	A_0 &=& \omega(r) ,
\label{sec1-40}\\
  A_\varphi &=& Q \cos \theta 
\label{sec1-50}
\end{eqnarray}
with the radial electric and magnetic fields
\begin{eqnarray}
	E_r &=& \frac{d \omega}{dr} = \frac{q a e^{-3\varphi}}{\Delta^2},
\label{sec1-60}\\
  H_r &=& \frac{Q}{a}.
\label{sec1-70}
\end{eqnarray}
But for the peripheral parts $\Delta (r) < 0$ that means that the time $t$ becomes as 
$5^{th}$ coordinate and $5^{th}$ coordinate $\chi$ becomes as time coordinate. 
\par 
At the tails according to equation \eqref{sec0-40}
\begin{eqnarray}
	\widetilde{\delta}(r) &=& \frac{\Delta(r)}{a(r)}\omega^2(r) - 
	\frac{a(r)}{\Delta(r) e^{2\phi(r)}},
\label{sec4-10}\\	
	\widetilde{A}_0(r) &=& \frac{\Delta(r)}{a(r) \widetilde{\delta}(r)} \omega(r) ,
\label{sec4-20}\\	
	\widetilde{A}_\varphi(r, \theta) &=& \frac{\Delta(r) \omega (r)}{a(r) 
	\widetilde{\delta}(r)} Q \cos \theta .
\label{sec4-30}
\end{eqnarray}

\section{Non-singularity of the gravitational flux tube metric}

The most important for the idea presented here is the regularity of the 
solution everywhere. 
Evidently that the metric \eqref{sec1:10} can have a singularity only at the points where 
$\Delta = 0$. In this section we would like to show that the gravitational flux tube solutions 
is non-singular at the points $\pm r_H$ where $\Delta(\pm r_H) = 0$. Let us  
investigate the solution near to the point $|r| \approx r_H$ where the metric functions 
have the following view 
\begin{eqnarray}
  a(r) &=& a_0 + a_1 \left ( r-r_H \right ) + 
  a_2 \left ( r-r_H \right )^2 + \cdots ,
    \label{sec1:51}\\
    \phi(r) & = & \phi_H + \phi_1 \left ( r-r_H \right ) + 
    \phi_2 \left ( r-r_H \right )^2 + \cdots ,
    \label{sec1:52}\\
    \Delta(r) & = & \Delta_1 \left( r - r_H \right) + 
    \Delta_1 \Delta_2 \left( r - r_H \right)^2 + \cdots .
    \label{sec1:54}
\end{eqnarray}
The substitution these functions in Einstein's equations 
\eqref{sec1:22}-\eqref{sec1:28} gives us the following coefficients 
\begin{eqnarray}
    \Delta_1 & = & \pm q e^{-2\phi_H},
    \quad (+) \quad \text{for} \quad r \rightarrow -r_H 
  \quad \text{and} 
  \quad (-) \quad \text{for} \quad r \rightarrow +r_H ,
    \label{sec1:56}\\
    \phi_2 &=& -\phi_1 \frac{a_1 + a_0\phi_1}{2a_0},
    \label{sec1:57}\\   
    \Delta_2 &=& \frac{-3a_0 \phi_1 + a_1}{2 a_0},
    \label{sec1:58}\\   
    a_2 &=& \frac{2 - a_1 \phi_1}{2}.
    \label{sec1:59}
\end{eqnarray}
In this case the electric field \eqref{sec1:30} has the following behaviour near to the 
points $r=\pm r_H$  
\begin{equation}
    \omega'(r) = \frac{a_0 e^{\phi_H}}{q} \frac{1}{\left( r - r_H \right)^2} + 
    \omega_1 + \mathcal{O}\left( r - r_H \right)
    \label{sec1:59a}
\end{equation}
where $\omega_1$ is some constant depending on $a_{0,1}, \phi_{1,2}, \Delta_{1,2}$. 
Then $\omega(r)$ is 
\begin{equation}
    \omega(r) = -\frac{a_0 e^{\phi_H}}{q} \frac{1}{\left( r - r_H \right)} + 
    \omega_0 + \mathcal{O}\left( r - r_H \right)
    \label{sec1:59b}
\end{equation}
where $\omega_0$ is some integration constant. The $G_{tt}$ metric component is 
\begin{equation}
    G_{tt} = \frac{a(r)}{\Delta(r)} - 
    \frac{\Delta(r) e^{2 \phi(r)}}{a(r)} \omega^2(r) = 
    -e^{2\phi_H} \frac{2qe^{-\phi_H} \omega_0 - a_1 - a_0 \phi_1}{q} + 
    \mathcal{O} \left( r - r_H \right).
    \label{sec1:59c}
\end{equation}
Finally the metric \eqref{sec1:10} has the following approximate behaviour 
near to $r = \pm r_H$ points \begin{equation}
\begin{split}
    &ds^2 = \left[ g_H +    \mathcal{O} \left( r - r_H \right) \right] dt^2 - 
    \mathcal{O} \left( r - r_H \right) 
    \left( d\chi + Q \cos\theta d \varphi \right)^2 - \\
    &\left[ e^{\phi_H} + \mathcal{O} \left( r - r_H \right) \right]
    dt \left( d\chi + Q \cos\theta d \varphi \right) - 
    dr^2 - \left[ a(r_H) + \mathcal{O} \left( r - r_H \right] \right) 
    \left( d\theta^2 + \sin^2 \theta d\varphi^2 \right) \approx \\
    &e^{\phi_H} dt \left( d\chi + Q \cos\theta d \varphi \right) - 
    dr^2 - a(r_H) \left( d\theta^2 + \sin^2 \theta d\varphi^2 \right)
\end{split} 
\label{sec1:59e}
\end{equation}
where 
$g_H = -e^{2\phi_H} \left (2qe^{-\phi_H} \omega_0 - a_1 - a_0 \phi_1 \right )/q$. 
It means that at the points $r = \pm r_H$ the metric \eqref{sec1:10} 
is non-singular one. 

\section{Asymptotical behaviour of the gravitational flux tube metric}

Let us consider $|r| \gg r_H$ parts of the solution. 
We search the asymptotical behavior of the gravitational flux tube metric as 
\begin{eqnarray}
    a(r) & = & r^2 + m_1 r + q_1 + \cdots ,
    \label{sec2:70}\\
    \Delta(r) & = & -\Delta_\infty r^2 + \Delta_\infty m_2 r + 
    \Delta_\infty q_2 + \cdots ,
    \label{sec2:80}\\
    \phi(r) & = & \phi_\infty + \frac{\phi_1}{r^2} + \cdots .
    \label{sec2:90}
\end{eqnarray}
The Einstein's equations give 
\begin{eqnarray}
    \phi_1 & = & -\frac{Q^2 \Delta_\infty e^{2 \phi_\infty}}{4},
    \label{sec2:100}\\
    q_1 & = & \frac{q^2 e^{-4 \phi_\infty} + 
    3Q^2 \Delta^3_\infty e^{2 \phi_\infty} - 
    \Delta^2_\infty m_2^2 - 2 \Delta^2_\infty m_1 m_2}{4 \Delta^2_\infty}, 
    \label{sec2:110}\\
    q_2 & = &  \frac{q^2 e^{-4 \phi_\infty} - 
    3Q^2 \Delta^3_\infty e^{2 \phi_\infty} - 
    \Delta^2_\infty m_2^2}{4 \Delta^2_\infty}.
    \label{sec2:120}
\end{eqnarray}
It shows us that at the infinity 
\begin{equation}
    \frac{\Delta(r) e^{2 \phi(r)}}{a(r)} \approx - \Delta_\infty e^{2\phi_\infty} 
    \left( 1 - \frac{m_1 + m_2}{r} - \frac{m_1^2 - q_1 - q_2 + 2\phi_1}{r^2} \right)
\label{sec2:130}
\end{equation}
The numerical investigations \cite{dzh2} give us arguments that 
$\Delta(r) e^{2 \phi(r)}/a(r) \rightarrow -1$ and consequently 
\begin{equation}
    \Delta_\infty = e^{-2\phi_\infty} .
    \label{sec2:135}
\end{equation}
After substitution in equations \eqref{sec2:100}-\eqref{sec2:120} we have 
\begin{eqnarray}
    \phi_1 & = & -\frac{Q^2}{4},
    \label{sec2:140}\\
    q_1 & = & \frac{q^2 + 3Q^2 - m_2^2 - 2 m_1 m_2}{4}, 
    \label{sec2:150}\\
    q_2 & = &  \frac{q^2 - 3Q^2 - m_2^2}{4}.
    \label{sec2:160}
\end{eqnarray}
Thus asymptotically we have 
\begin{equation}
	\frac{d \omega}{dr} = \frac{q a}{\Delta e^{2\phi}} \frac{e^{-\phi}}{\Delta} 
	\rightarrow \frac{q e^{\phi_\infty}}{r^2}.
\label{sec5-10}
\end{equation}
Therefore
\begin{equation}
	\omega = - \frac{q e^{\phi_\infty}}{r}.
\label{sec5-20}
\end{equation}
According to equation \eqref{sec0-40}
\begin{eqnarray}
	\widetilde{\delta} &=& \frac{\Delta}{a}\omega^2 - \frac{a}{\Delta e^{2\phi}} 
	\rightarrow 1,
\label{sec5-30}\\	
	\widetilde{A}_0 &\rightarrow & \frac{q e^{-\phi_\infty}}{r},
\label{sec5-40}\\	
	\widetilde{A}_\varphi &\rightarrow& \frac{q Q}{r} e^{-\phi_\infty} \cos \theta
\label{sec5-50}
\end{eqnarray}
and asymptotically the metric is 
\begin{equation}
	ds^2 \approx 
	-\left(
		d \widetilde{t} + \frac{q}{r} d \chi + \frac{qQ}{r} \cos \theta d \varphi
	\right)^2 + 
	\left(
		d\chi + Q\cos \theta d \varphi
	\right)^2 - 
	dr^2 - r^2 
	\left(
		d \theta^2 + \sin^2 \theta d \varphi ^2
	\right)
\label{sec5-60}
\end{equation}
where $\widetilde{t} = e^{\phi_\infty}t$. We see that asymptotically there are 
the radial electric and magnetic fields 
\begin{equation}
	\left| E_r \right| \approx \frac{q}{r^2}, \quad 
	\left| H_r \right| \approx \frac{q Q}{r^3}.
\label{sec5-70}
\end{equation}
Remarkably that \textit{the magnetic field is suppressed} in comparison with 
the electric field. 
\par 
The 4D part of the metric \eqref{sec5-60} is 
\begin{equation}
	ds^2_{(4)} \approx 
	Q^2 \left(
		\frac{d\chi}{Q} + \cos \theta d \varphi
	\right)^2 - r^2 
	\left(
		d \theta^2 + \sin^2 \theta d \varphi ^2
	\right) - dr^2 
\label{sec5-80}
\end{equation}
The first two terms give the Lorentzian metric on the deformed Hopf bundle 
$S^3 / S^1 \rightarrow S^2$ where the total space $S^3$ is spanned on coordinates 
$\chi, \theta, \varphi$; the fiber $S^1$ on $\chi$ coordinate and the base 
$S^2$ on $\theta, \varphi$ coordinates. On the principal Hopf bundle the angle 
$0 \leq \chi \leq 4 \pi$ consequently the time $\chi$ is closed one and topologically 
non-trivial in the sense that the spacetime is not the direct product: 
\textit{spacetime $\neq$ time $\times$ space}. Physically it means that we can not 
introduce a global time but only a local one. The situation is similar to G\"odel's 
and Taub-NUT's metrics. In order to avoid closed time one can consider the associated 
Hopf bundle with the fibers $R^1 = u(1) = \mathrm{Lie} \{ U(1) \}$. In this case 
the time $\chi$ becomes infinite but nevertheless the bundle remains non-trivial 
and as before we can not introduce a global time.
\par
Finally we would like to note that the off-diagonality of the metrics 
\eqref{sec0-5} \eqref{sec0-10} is very important that was emphasized 
in Ref. \cite{canschmidt}.

\section{A duality between magnetic field and an angular momentum density}

Another interesting peculiarity of the gravitational flux tube spacetime is that 
by $|r| > r_h$ the metric has the 4D off-diagonal metric component 
\begin{equation}
	g_{\chi \varphi} = \frac{1}{\widetilde{\Delta}} Q \cos \theta \: d \chi d \varphi
\label{sec6-10}
\end{equation}
where $\chi$ is the time-like coordinate. It shows us that like to Kerr and G\"odel metrics
in this spacetime there is a rotation. 
\par 
In order to understand what kind of the rotation there is here let us consider the 
electromagnetic field. From equation \eqref{sec0-40} we see that by $|r| > r_H$ 
we have the following electromagnetic potential 
\begin{eqnarray}
	\widetilde{A}_0 &=& - \frac{\Delta (r)}{a(r) \widetilde{\Delta}(r)} \omega(r) ,
\label{sec6-20}\\
	\widetilde{A}_\varphi &=& - \frac{\Delta (r) \omega (r)}{a(r) \widetilde{\Delta}(r)} 
	Q \cos \theta .
\label{sec6-30}
\end{eqnarray}
Consequently we have the following non-zero components of the tensor of the 
electromagnetic field  
\begin{eqnarray}
	\widetilde{F}_{r \chi} &=& - \frac{d}{dr}
	\left(
		\frac{\Delta (r)}{a(r) \widetilde{\Delta}(r)} \omega(r) 
	\right) \neq 0 ,
\label{sec6-40}\\
	\widetilde{F}_{\theta \phi} &=& 
		\frac{\Delta (r) \omega(r)}{a(r) \widetilde{\Delta}(r)} Q \sin \theta 
	\neq 0 ,
\label{sec6-50}\\
	\widetilde{F}_{r \phi} &=& 
	Q \cos \theta \frac{d}{dr}
	\left(
		\frac{-\Delta (r) \omega(r)}{a(r) \widetilde{\Delta}(r)} 
	\right) \neq 0 ,
\label{sec6-60}
\end{eqnarray}
The definition of the tensor of the angular momentum density for the electromagnetic 
field is 
\begin{equation}
	M^{\mu \nu} = x^\mu T^{\nu 0} - x^\nu T^{\mu 0}
\label{sec6-70}
\end{equation}
where 
\begin{equation}
	T^{\mu \nu} = F^{\mu \alpha} F^\nu_{\  \alpha} - 
	\frac{1}{4}g^{\mu \nu} F_{\alpha\beta} F^{\alpha\beta}
\label{sec6-80}
\end{equation}
is the energy - momentum tensor for the electromagnetic field. The substitution 
electromagnetic field \eqref{sec6-40}-\eqref{sec6-60} into definitions \eqref{sec6-70} 
and \eqref{sec6-80} gives us that 
\begin{equation}
	M^{r \theta}, M^{r \varphi} \neq 0 .
\label{sec6-90}
\end{equation}
In vector notation the angular momentum is defined as 
\begin{equation}
	M_i = \epsilon_{ijk} M^{jk} .
\label{sec6-100}
\end{equation}
According to \eqref{sec6-90} 
\begin{equation}
	M_\varphi , M_\theta \neq 0 .
\label{sec6-110}
\end{equation}
Thus the gravitational flux tube solutions have the rotation term \eqref{sec6-10} 
in the metric connected with the angular momentum density of the electromagnetic 
field \eqref{sec6-110}. 
\par 
It is necessary to deduce why the metric \eqref{sec1:10} has not the form of the 
Kerr metric for the rotating black hole. The reason is that the $\chi$ coordinate 
is topologically non-trivial. At the center $(r = 0)$ the metric is 
\begin{equation}
	ds^2 = \frac{a(0)}{\Delta(0)} dt^2 - dr^2 - 
	\left[
		Q^2 \frac{\Delta(0)}{a(0)} e^{2 \phi(0)} 
			\left(
				\frac{d \chi}{Q} + \cos \theta d \phi
			\right)^2 + a(0) 
			\left(
				d \theta^2 + \sin \theta d \varphi^2
			\right)
	\right] 
\label{sec6-130}
\end{equation}
as $\omega(0) = 0$. We can choose the scale of time in such a way that $a(0)/\Delta(0) = 1$. 
One can see that the last term 
\begin{equation}
	dl^2 = 	Q^2 e^{2 \phi_0} 
			\left(
				\frac{d \chi}{Q} + \cos \theta d \varphi
			\right)^2 + a_0 
			\left(
				d \theta^2 + \sin \theta d \phi^2
			\right) .
\label{sec6-140}
\end{equation}
again is the metric on the deformed $S^3-$sphere presented as the Hopf bundle: 
$S^3 / S^1 \rightarrow S^2$. The coordinate $\chi$ is directed 
along the fibers. Thus one can say that the space-like coordinate $\chi$ on the throat 
and the time-like coordinate $\chi$ on the tails of the gravitational 
flux tube spacetime are non-trivial. This non-triviality in combination with the 
non-zero angular momentun density of the electromagnetic field show us that on the tails 
of gravitational flux tube solutions there is \textit{a rotation connected with the 
magnetic field on the throat.}

\section{Conclusions and discussion}

In this notice it is shown that:
\begin{itemize}
	\item 
	The regular gravitational flux tube metric has two different kind of metric 
	signature: the first type is $\{+,-,-,-,-,\}$ on the throat by $|r| < r_H$, 
	the second one is $\{-,-,-,-,+,\}$ on the tails by $|r| > r_H$. On the throat 
	there are the electric and magnetic fields.
	\item 
	On the tails the magnetic field decreases faster the electric field, although 
	on the throat the fields are almost equal. 
	\item 
	On the tails there is a rotation connected with the magnetic field on the 
	throat, i.e. a rotation on an external universe and the magnetic field on the 
	throat are dual each other. 
	\item 
	It is very important that the gravitational flux tube solutions are the 
	vacuum solutions of the 5D Einstein's equations and consequently does not 
	depend on the properties of any matter sort.
	\item 
	The time direction on the tails is topologically non-trivial. Probably 
	it is like to the time in the G\"odel solution.
	\item 
	Probably the most interesting case is when the length of $\chi$ coordinate 
	is in the Planck region and $\delta_q = 1 - Q/q \ll 1$. In this case this 
	dimension is invisible and effectively we have two Euclidean spacetime 
	connected with super-thin and super-long flux tube 
	(namely, $\Delta-$string \cite{dzh2}).
\end{itemize}
The most interesting question arising here is: one can extend these results to 
the topologically trivial time when $spacetime = space \times time$ ? In this case 
the gravitational flux tube solutions can be a geometrical model of electric 
charge with the explanation why the magnetic charge is unobservable in the 
nature. 
\par 
It is necessary to note that although the metric \eqref{sec1:10} at the 
tails is spherically symmetric one but this solution is not listed 
in review \cite{cveticyoum} as the presented solution has a rotation 
connected with the magnetic field and consequently the off-diagonal 4D metric 
component. 

\section{Acknowledgment}
I am very grateful to the ISTC grant KR-677 for the financial support.

\end{document}